\documentclass[12pt]{article}
\usepackage{graphicx}

\newcommand\pubnumber{Article 41 in eConf C1304143}
\newcommand\pubdate{\today}

\def\ste{Solar-Terrestrial Environment Laboratory, Nagoya University, Japan}
\def\nagoya{Division of Paticle and Astronomical Science, Graduate School of Science, Nagoya University, Japan}
\def\agu{Department of Physics and Mathematics, Aoyama Gakuin University, Japan}
\def\jaxa{Institute of Space and Astronautical Science (ISAS), Japan Aerospace Exploration Agency (JAXA), Japan}
\def\waseda{Research Institute for Science and Engineering, Waseda University, Japan}
\def\lsu{Department of Physics and Astronomy, Lousiana State University, USA}
\def\cnr{Istituto di Fisica Applicata ``Nello Carrara'' (IFAC), CNR, Italy}

\def\Title#1{\begin{center} {\Large #1 } \end{center}}
\def\Author#1{\begin{center}{ \sc #1} \end{center}}
\def\Address#1{\begin{center}{ \it #1} \end{center}}

\newcommand\pubblock{\rightline{\begin{tabular}{l} \pubnumber\\
         \pubdate  \end{tabular}}}
\newenvironment{Abstract}{\begin{quotation}  }{\end{quotation}}
\newenvironment{Presented}{\begin{quotation} \begin{center}
             PRESENTED AT\end{center}\bigskip
      \begin{center}\begin{large}}{\end{large}\end{center} \end{quotation}}
\def\Acknowledgements{\bigskip  \bigskip \begin{center} \begin{large}
             \bf ACKNOWLEDGEMENTS \end{large}\end{center}}

\def\beq{\begin{equation}}
\def\eeq#1{\label{#1}\end{equation}}
\def\eeqn{\end{equation}}

\def\beqa{\begin{eqnarray}}
\def\eeqa#1{\label{#1}\end{eqnarray}}
\def\eeqan{\end{eqnarray}}

\let\bar=\overbar

\def\Dslash{\not{\hbox{\kern-4pt $D$}}}
\def\dslash{\not{\hbox{\kern-2pt $\del$}}}

\def\msb{{\bar{\ssstyle M \kern -1pt S}}}

\begin{document}
\begin{titlepage}
\pubblock

\vfill
\Title{The CALET Gamma-ray Burst Monitor (CGBM)}
\vfill
\Author{Kazutaka Yamaoka}
\Address{\ste}
\Address{\nagoya}
\Author{Atsumasa Yoshida, Takanori Sakamoto, Ichiro Takahashi,  Takumi Hara, Tatsuma Yamamoto, Yuta Kawakubo, Ryota Inoue, Shunsuke Terazawa, Rie Fujioka, Kazumasa Senuma}
\Address{\agu}
\Author{Satoshi Nakahira, Hiroshi Tomida, Shiro Ueno}
\Address{\jaxa}
\Author{Shoji Torii} 
\Address{\waseda}
\Author{Michael L. Cherry}
\Address{\lsu}
\Author{Sergio Ricciarini}
\Address{\cnr}
\Author{and the CALET collaboration}
\vfill
\begin{Abstract}
The CALET Gamma-ray Burst Monitor (CGBM) is the secondary scientific instrument of the CALET mission on the International Space Station (ISS), which is scheduled for launch by H-IIB/HTV in 2014. The CGBM provides a broadband energy coverage from 7 keV to 20 MeV, and simultaneous observations with the primary instrument Calorimeter (CAL) in the GeV - TeV gamma-ray range and Advanced Star Camera (ASC) in the optical for gamma-ray bursts (GRBs) and other X-gamma-ray transients. The CGBM consists of two kinds of scintillators: two LaBr$_3$(Ce) (7 keV - 1 MeV) and one BGO (100 keV - 20 MeV) each read by a single photomultiplier. The LaBr$_3$(Ce) crystal, used in space for the first time here for celestial gamma-ray observations, enables GRB observations over a broad energy range from low energy X-ray emissions to gamma rays. The detector performance and structures have been verified using the bread-board model (BBM) via vibration and thermal vacuum tests. The CALET is currently in the development phase of the proto-flight model (PFM) and the pre-flight calibration of the CGBM is planned for August 2013. In this paper, we report on the current status and expected performance of CALET for GRB observations.
\end{Abstract}
\vfill
\begin{Presented}
GRB 2013 \\
the Seventh Huntsville Gamma-Ray Burst Symposium \\
Nashville, Tennessee, 14--18 April 2013
\end{Presented}
\vfill
\end{titlepage}
\def\thefootnote{\fnsymbol{footnote}}
\setcounter{footnote}{0}

\section{Introduction}

The gamma-ray burst (GRB) is one of the most energetic events in the universe, 
 with a radiated energy release of $>$10$^{51}$ erg.  For prompt emissions, most of the energies 
 are emitted in an energy range from soft X-rays to GeV gamma-rays. 
 It has been widely believed that electromagnetic radiation in the 10 keV--10 MeV range 
 are probably due to optically thin synchrotron emission from accelerated electrons 
 in relativistic jets. However, Fermi LAT observations have revealed a common feature 
 that there is an additional hard, delayed and long-lived component in GeV gamma-rays for both short 
 and long-duration GRBs, e.g. \cite{fermi_grb1} and \cite{fermi_grb2}. Leptonic and hadronic 
 models are still debated. On the other hand, Beppo-SAX and HETE2 revealed a presence of X-ray flashes 
 (XRF) whose emissions are dominant in X-rays rather than gamma-rays. Judging from spectral and timing 
 properties, the XRFs are considered to belong the same class as GRBs \cite{xrf}. Thus, to understand 
 radiation mechanisms of GRBs, the importance of wide-band X-and gamma-ray observations 
 of GRBs are growing. 

\begin{figure}[!t]
  \centering
  \includegraphics[width=0.55\textwidth]{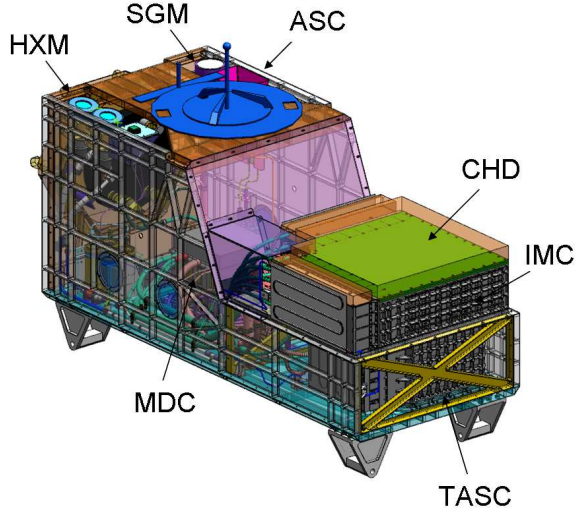}
  \caption{Overall picture of the CALET mission payload on the ISS. The CGBM, consisting of HXM and SGM, is attached on the top of the CALET for both sides.}
  \label{fig1}
 \end{figure}

The Calorimetric Electron Telescope (CALET \cite{calet}) is designed primarily for observation 
 of high energy electrons and gamma-rays in the GeV -- TeV range. It will be launched by the
 H-IIB/H-II Transfer Vehicle (HTV), and attached to Exposed Facility of the Japanese Experimental 
  Module (JEM) on the International Space Station (ISS) in 2014. The overview of 
 the CALET mission payload is shown in Figure \ref{fig1}. The primary instrument, 
 calorimeter (CAL), can observe a GRB in the GeV -- TeV gamma-ray range. 
 In order to enhance the capability for GRB observations with CAL, we have added a 
 gamma-ray burst monitor (CGBM), which is sensitive to X-rays to soft gamma-rays up to 20 MeV. 
 We also actively utilize the Advanced Star Camera (ASC), which is attached for an accurate localization of 
 gamma-ray sources, to observe an optical flash during the GRB prompt emissions \cite{optflash} 
 with a limiting magnitude of about 9 in optical (see Table \ref{tab1}). 
 By investigating very broadband coverage from optical, X-rays to GeV--TeV gamma-rays,
  we will clarify the following issues: 1) radiation mechanisms for GRB prompt emissions including XRFs, 
 2) search for a coincidence to a gravitational wave (GW) event as expected to start operations of 
 upgraded GW detectors in later 2010s, and 3) presence for absorption features as seen 
 in the Ginga/GBD energy spectrum of GRB 880205 \cite{ginga_abs}. 
 Furthermore, we emphasize that we can also perform simultaneous 
 X-ray observations with MAXI \cite{maxi} attached to the same facility JEM-Kibo on the ISS. 

\begin{table}[htbp]
\begin{center}
\caption{GRB Observations with CALET.}\label{tab1}
\begin{tabular}{lccc}\\\hline\hline
                        &   CAL                 & CGBM              & ASC \\\hline    
Energy (Wavelength)     &  a few GeV--10 TeV & 7 keV--20 MeV       & 300$\sim$800 nm  \\
Effective area          &  $\sim$600 cm$^2$    &  68 cm$^2$(HXM)   &  --- \\
                        &                      &    82 cm$^2$(SGM) &       \\  
Field of view           &  $\sim$2 str.          &  $\sim$3 str.(HXM)  & 18.4$^{\circ}\times$13.4$^{\circ}$ \\
                        &                        &  4$\pi$ str.(SGM)   &    \\
Angular resolution      &  2.5$^{\circ}$ @1 GeV  &   No capabilities & ---  \\
                        &  0.35$^{\circ}$ @10 GeV &                  &      \\ 
Time resolution         &   62.5 $\mu$s         &   62.5 $\mu$s    & 1/16$\sim$4 sec   \\
Deadtime per event      &   1.8 ms              &   40 $\mu$s  & ---    \\\hline
\end{tabular}
\end{center}
\end{table}

\section{Instrumentation}
\subsection{Overview}

The CGBM is a secondary scientific instrument which supports a capability for GRB observations
 by the CAL. It observes a lower energy range which the CAL can not cover 
 and is aimed to obtain GRB spectra over a wide range of X-rays 
 to gamma-rays. In addition to GRB observations, the CGBM will carry out all-sky observations of 
 various gamma-ray transients: soft gamma repeaters, solar flares, terrestrial gamma-ray flashes, 
 and X-ray binaries.  It is composed of three sensors and an electronics box (E-box) which processes 
the signal of sensors. The outputted digitized data from the E-box are formatted by the Mission Data
Controller (MDC), and sent as a telemetry to the ground station at NASA.

\subsection{Sensors}

The CGBM consists of two types of detectors, a Hard X-ray Monitor (HXM) and a Soft Gamma-ray
Monitor (SGM), to cover a wide energy range of 7 keV to 20 MeV by its combination. 
 As for the HXM, we utilize a novel scintillator
 LaBr$_3$(Ce) with excellent performance in terms of light yield, energy
 resolution, and time response in comparison with NaI(Tl). 
 However, it is used in space for the first time here for celestial gamma-ray observations. 
 Hence, we have been verifying that the performance will not be degraded 
   through proton and gamma-ray irradiation tests.   
 Each of the two LaBr$_3$(Ce) crystals is configured as two
 cylinders, the front cylinder 66.0 mm diameter and 6.35 mm thick and the rear
 cylinder 78.7 mm diameter and 6.35 mm thick. The beryllium entrance window
 with a 410 mm thickness is used for soft X-ray detection below 10 keV. 
 For the SGM, we utilize the BGO scintillator which 
 has a high stopping power for gamma-rays due to its large density ($\rho$=7.13 gcm$^{-3}$) 
 and effective atomic number ($Z_{\rm eff}$=74). The BGO crystal has a cylindrical shape 
 with 102 mm diameter and 76 mm thickness. 
The LaBr$_3$(Ce) and BGO crystal units are provided by 
 Saint Gobain Crystals and OKEN Co., Ltd. respectively. 

\begin{figure}[!t]
  \centering
  \includegraphics[width=0.55\textwidth]{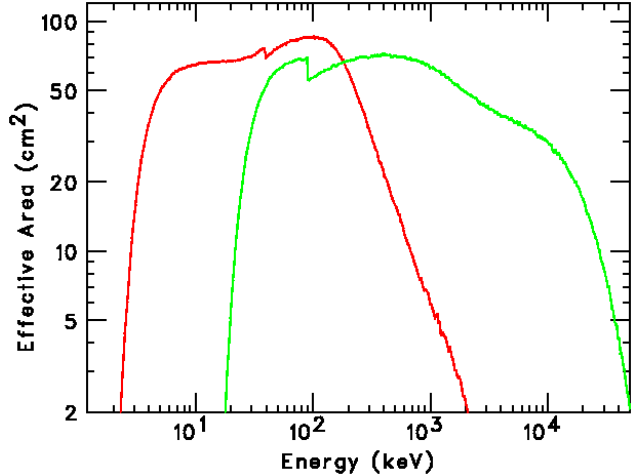}
  \caption{Calculated on-axis effective area of the CGBM HXM (shown in red) and SGM (green). The HXM is sensitive to soft X-rays below 10 keV.}
  \label{fig2}
 \end{figure}

The CGBM sensors are in total three detectors, two identical HXMs and one SGM. Figure \ref{fig2} 
 shows the HXM and SGM effective area as a function of energy. The unique feature for the HXM 
 is a sensitivity to soft X-rays below 10 keV. The HXM covers a lower energy range of 7--1000 keV, 
 while the SGM covers a higher energy of 100 keV--20 MeV. 
We aim at operating both detectors at a lower energy threshold. 
Each sensor mainly contains a scintillation crystal, a photo-multiplier tube 
 (PMT), a high voltage divider, and a charge sensitive amplifier (CSA). The vibration-proof model 
 of Hamamatsu PMT R6232--05 (2.4 inch diameter) and  R6233--20 (3 inch diameter) with a high 
 quantum efficiency are used for the HXM and the SGM respectively. 
 All the components are installed in an aluminum housing. 
 The field of view for the HXM is 
 limited by the light collimator within about 58 degrees from the zenith to reduce possible
  contamination from the cosmic X-ray background and bright X-ray sources.  
 In this configuration, we expect a GRB detection rate of about 30--40 per year by the HXM. 

\subsection{GBM Electronics-box (E-box)}

\begin{figure}[!t]
  \centering
  \includegraphics[width=0.7\textwidth]{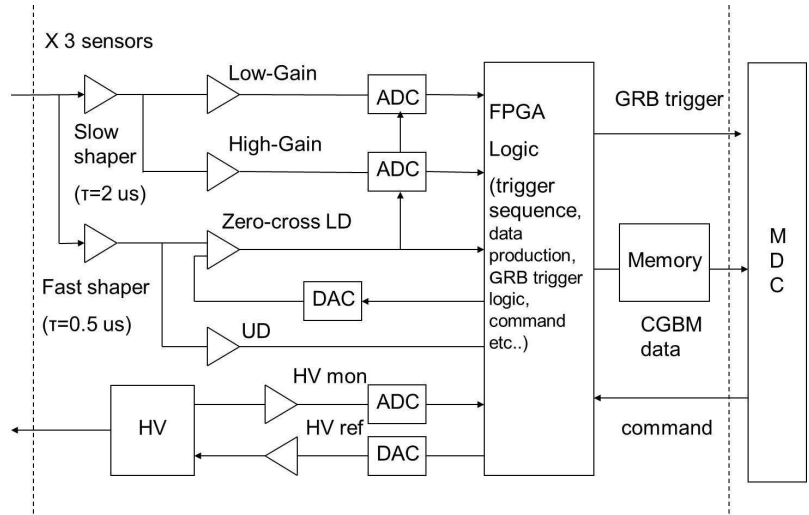}
  \caption{Signal processing of the CGBM sensors in the GBM E-box. The analog and digital circuits, ADCs, FPGAs, 10 Mbyte ring-buffer memory, and high voltages are included in the E-box.}
  \label{fig3}
 \end{figure}

The CGBM E-box processes three signals from the CSA 
 in the three sensors. It is located apart from both sensors (HXM and SGM)
 and MDC. The block diagram of signal processing in the E-box
 is shown in Figure \ref{fig3}. The HXM and SGM signal processing 
 are the same. To ensure a wide dynamic range of a factor of $\sim$1000, the signals 
 are divided into two amplifiers with different gains (high and low), 
 and read out by different sample and hold 16-bit ADCs. The signals from the CSA are
 bipolar-shaped with a CR-RC$^2$-CR filter where the time constant 
 $\tau$ is 2 $\mu$s. The ADC conversion timing is produced by 
 a zero-crossing lower discriminator after the shaping with a faster time
 constant ($\tau$=0.5 $\mu$s). 
The dead-time per event is set at 40 $\mu$s. 
 An upper discriminator (UD) is also present in the circuit with the purpose of  
 avoiding malfunction in case of large energy-deposit signals.  The digitized 
 data from ADC and comparators are sent to Field Programmable Gate 
 Arrays (FPGAs). The FPGAs implement a trigger sequence, scalers from LDs and UDs, 
 data formatting, a GRB trigger logic, command decoders and so on. 
 The three high voltages S9099, provided by SITAEL, are installed in the E-box 
 for each sensor, and are controlled individually with 8 bit in the 0--+1250 V range.  

 The CGBM E-box produces two types of the scientific data (see Table \ref{tab3}). 
 One is continuous monitoring data 
  which is always outputted to the telemetry, independent of the GRB trigger status.  
 The monitor data has two kinds of histogram data: Time History (TH) data with a 1/8 sec time 
 resolution and 8 energy channels, and Pulse Height (PH) data with a 4 sec time resolution
 and 512 energy channels. 
 The other is an event-by-event data which are available 
 only when a GRB triggers the CGBM. It has fine information of an arrival time with 
 62.5 $\mu$s time resolution and an energy with 4096 channels for each high and low gain. 

 On-board trigger system is realized in the FPGA of the E-box. The following 
 criterion for the GRB trigger is installed in the hardware:
\begin{equation}
N_{\rm tot}-\frac{N_{\rm BG}}{\Delta t_{\rm BG}}\Delta t>\sigma\sqrt{\frac{N_{\rm BG}}{\Delta t_{\rm BG}}\Delta t}
\end{equation} 
where $N_{\rm tot}$ is the GRB plus background counts during the GRB judgment 
 time ($\Delta t$=1/4, 1/2, 1 and 4 seconds), $N_{\rm BG}$ is the background counts 
 during the background integration time ($\Delta t_{\rm BG}$), and $\sigma$ is the significance 
 level.  If the source counts exceeds $\sigma$ times the statistical fluctuation of the background counts,  
 we can regard it as a GRB event. 
This logic is very simple, and has been successfully applied to the previous GRB instruments such 
 as Ginga/GBD \cite{gbd} and Suzaku/WAM \cite{wam}. When a GRB triggers the CGBM, 
 the energy threshold for CAL will be set lower from 10 GeV to a few GeV, and 
 two ASC optical images will be taken.  

\begin{table}[htbp]
\begin{center}
\caption{CGBM Data Types and Contents}\label{tab3}
\begin{tabular}{lccc} \hline\hline\\[-6pt]
Data Type        & \multicolumn{2}{c}{Monitor Data}   &  Event Data  \\   
                 &  TH  & PH      &       \\   \hline
Energy channel   & 8    & 512 & 4096 (for both gains) \\
Time resolution  & 1/8 s & 4 s  & 62.5 $\mu$s   \\
Time coverage    & \multicolumn{2}{c}{anytime} & $\sim$1.5$\times$10$^6$ events\\
                 & \multicolumn{2}{c}{}        & around the trigger time \\ \hline
\end{tabular}
\end{center}
\end{table}

\section{Bread Board Model of Sensors and E-box}

The CALET is currently in the development and verification phase of the proto-flight model (PFM), 
 and the CGBM PFM will be completed till this July.
 Before starting productions of the PFM, we have verified mechanical designs and performance 
 using the bread board model (BBM). The BBM was prepared only for SGM since the HXM has 
 similar structure to the SGM. 

\begin{figure}[!t]
  \centering
  \includegraphics[width=0.39\textwidth]{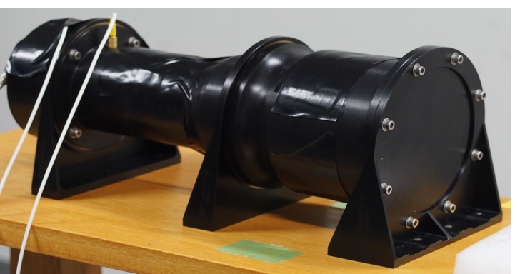}
  \includegraphics[width=0.27\textwidth]{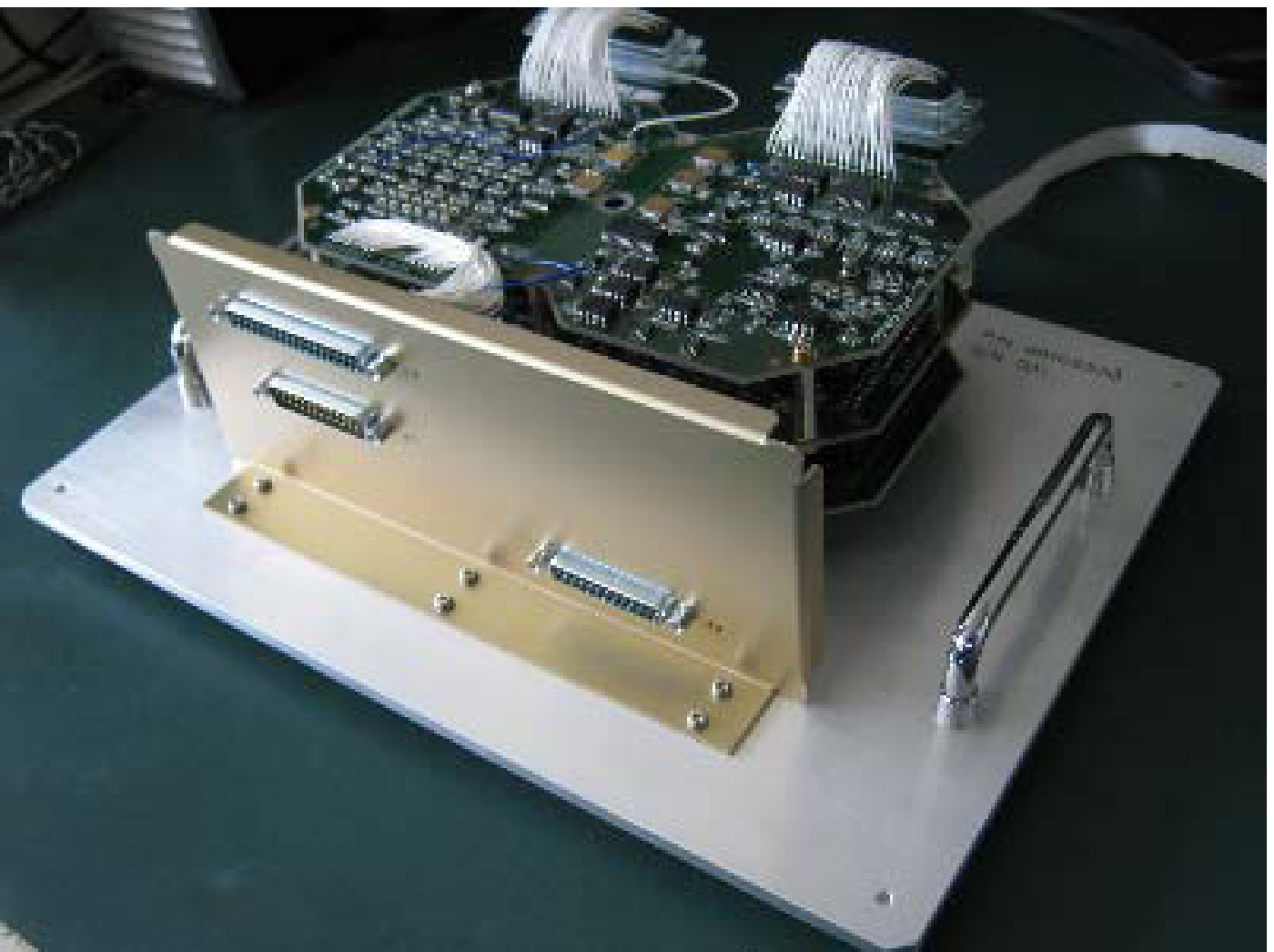}
  \caption{A picture of the SGM BBM sensor (Left) and GBM E-box (Right).}
  \label{fig4}
 \end{figure}

 A picture of the SGM BBM sensor is shown in left panel of Figure \ref{fig4}. 
 The BGO crystal unit was produced by OKEN Co., Ltd and installed in an aluminum housing painted black 
 on the outside and attached to a PMT R6233--20 by optical compound KE1051J (Shin-etsu Chemical Co., Ltd).  
 The vibration and thermal-vacuum test were done at facilities (IMV Corporation and ISAS/JAXA respectively)
 on this January. After checking resonance characteristics with the modal survey,
  a random vibration level of 19.8 Grms in the frequency range of 20--2000 Hz was set  
 during 120 seconds for the 3 axes. No critical damages on the sensors were seen during vibration tests.  
 For the thermal-vacuum test, we imposed four temperature cycles to the SGM BBM with a 
 high voltage turned on. The temperature range was --30 to 45 $^{\circ}$C with a variation rate 
 of 12 $^{\circ}$C per hour. We found that no clear discharge events were seen and the spectral 
 performance was unchanged during the thermal-vacuum test. Finally we disassembled the BBM, then 
 verified by eye that no cracks or anomaly was present in the BGO crystal unit. 

\begin{figure}[!t]
  \centering
  \includegraphics[width=0.6\textwidth]{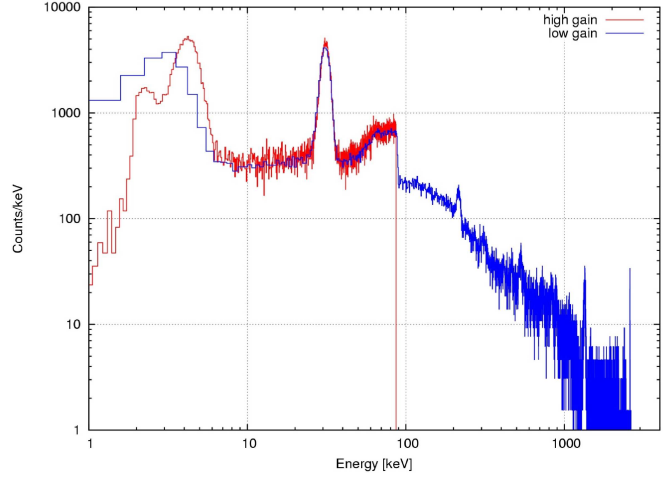}
  \caption{$^{232}$Th broadband spectrum taken by proto-type LaBr$_3$ detector with 220 $\mu$m-thickness beryllium window and the BBM of the GBM E-box. The spectrum from high and low gain is shown in red and blue respectively. We can clearly see several lines from 4.5 keV (Ba L$_{\rm x}$) and 32 keV (Ba K$_{\rm x}$) in low energies to 1436 keV ($^{138}$La) and 2614 keV ($^{208}$Tl) in high energies. }
  \label{fig5}
 \end{figure}

The BBM of the pre-amplifier and E-box was also fabricated.  
A picture of the GBM E-box is shown in right panel of Figure \ref{fig4}. 
 We measured a broadband spectrum with a proto-type 3-inch  LaBr$_3$(Ce) detector irradiated 
with a $^{232}$Th source. As can be seen from Figure \ref{fig5}, in addition to the 2.6 MeV 
 gamma-ray line from $^{208}$Tl, 
 we clearly see 4.5 keV L$_{\rm x}$ and 32 keV K$_{\rm x}$ lines from $^{138}$Ba when the radio isotope 
 $^{138}$La involved in LaBr$_3$ decays via an electron capture. Therefore a wide dynamic range of the 
 E-box has been verified. The high counting rate tolerance was also 
 verified up to a few 10 kHz which is expected for previous intense GRBs. 

\Acknowledgements

This research has been supported in part by a Grant-in-Aid for Scientific Research (24684015 KY) of the Ministry of Education, Culture, Sports, Science and Technology (MEXT).

\end{document}